\documentclass[12pt]{article}
\usepackage{cite,graphicx,epsfig}
\newcommand{\ba}{\begin{array}}
\newcommand{\ea}{\end{array}}
\newcommand{\be}{\begin{equation}}
\newcommand{\ee}{\end{equation}}
\newcommand{\bea}{\begin{eqnarray}}
\newcommand{\eea}{\end{eqnarray}}

\newcommand{\bin}[2]{\left( \begin{tabular}{c} $#1$ \\ $#2$ \end{tabular} \right)}

\def\IB{\relax\hbox{$\inbar\kern-.3em{\rm B}$}}
\def\IC{\relax\hbox{$\inbar\kern-.3em{\rm C}$}}
\def\ID{\relax\hbox{$\inbar\kern-.3em{\rm D}$}}
\def\IE{\relax\hbox{$\inbar\kern-.3em{\rm E}$}}
\def\IF{\relax\hbox{$\inbar\kern-.3em{\rm F}$}}
\def\IG{\relax\hbox{$\inbar\kern-.3em{\rm G}$}}
\def\IGa{\relax\hbox{${\rm I}\kern-.18em\Gamma$}}
\def\IH{\relax{\rm I\kern-.18em H}}
\def\IK{\relax{\rm I\kern-.18em K}}
\def\IL{\relax{\rm I\kern-.18em L}}
\def\IP{\relax{\rm I\kern-.18em P}}
\def\IR{\relax{\rm I\kern-.18em R}}
\def\IZ{\relax{\rm Z\kern-.5em Z}}

\def\half{\frac{1}{2}}

\def\f{\frac}

\def\TL{Temperley-Lieb }

\begin{document}
\begin{titlepage}

\vskip 2cm

\begin{center}
{\LARGE Structure of the two-boundary XXZ model with non-diagonal boundary terms}
\vskip 1 cm

{\large A. Nichols\footnote{nichols@sissa.it}}

\begin{center}
{\em International School for Advanced Studies,\\
Via Beirut 1, 34100 Trieste, Italy.}\\
{\em Istituto Nazionale di Fisica Nucleare (INFN), \\
Sezione di Trieste.} 
\end{center}

\vskip 1 cm

\begin{abstract}
We study the integrable XXZ model with general non-diagonal boundary terms at both ends. The Hamiltonian is considered in terms of a two boundary extension of the  Temperley-Lieb algebra.

We use a basis that diagonalizes a conserved charge in the one-boundary case. The action of the second boundary generator on this space is computed. For the $L$-site chain and generic values of the parameters we have an irreducible space of dimension $2^L$. However at certain critical points there exists a smaller irreducible subspace that is invariant under the action of all the bulk and boundary generators. These are precisely the points at which Bethe Ansatz equations have been formulated. We compute the dimension of the invariant subspace at each critical point and show that it agrees with the splitting of eigenvalues, found numerically, between the two Bethe Ansatz equations.
\end{abstract}

\end{center}

\end{titlepage}

\newpage
\section{Introduction}
\renewcommand{\theequation}{\arabic{section}.\arabic{equation}}
\setcounter{equation}{0}  
The spin-$\half$ XXZ model with general non-diagonal boundaries has been the subject of recent interest. Although this is well known to be integrable \cite{deVega:1992zd} many problems have been encountered in the formulation of Bethe Ansatz equations.

For very special diagonal boundary terms the XXZ model has an $SU_q(2)$ quantum group symmetry \cite{Pasquier:1989kd}. From an algebraic point of view this is the simplest type of boundary term and the model can be written in terms of the Temperley-Lieb algebra \cite{Temperley:1971iq}. 

The addition of a single boundary term to the $SU_q(2)$ chain can be described
using the one-boundary Temperley-Lieb (1BTL)
algebra \cite{Martin:1992td,Martin:1993jk,MartinWoodcockI,MartinWoodcockII}. Although the integrable Hamiltonian involves three parameters only two of these appear in the 1BTL algebra. It is the algebraic parameters which control the structure of the lattice theory.

As shown in \cite{Nichols:2004fb} this general one-boundary 
Hamiltonian has exactly the same spectrum as the XXZ Hamiltonian with purely diagonal
boundary terms. The Bethe Ansatz for the XXZ chain with purely diagonal boundary terms is especially simple due to the presence of an obvious conserved charge and Bethe reference states \cite{Alcaraz:1987uk}. 

The situation of non-diagonal boundary terms at both ends of the chain
is considerably more complicated. One now has five boundary
parameters. As noticed in \cite{deGier:2003iu} the Hamiltonian
can be written in terms of the generators of the two-boundary Temperley-Lieb (2BTL) 
algebra \cite{JanReview}. This algebra depends on three
boundary parameters only with the other two parameters of the problem entering as coefficients in the integrable Hamiltonian.

The formulation of Bethe Ansatz equations for the general two-boundary system has attracted recent attention. Unlike the diagonal case there is no obvious Bethe reference state. For the free fermion point the spectrum and wavefunctions can be found \cite{BilsteinI}. However away from this point, apart from some special cases \cite{Murgan:2005rz,Murgan:2005bp,Yang:2005ce}, the Bethe
ansatz equations have only been obtained in the case in which the parameters satisfy an additional constraint \cite{ChineseGuys,Nepomechie:2002xy,Nepomechie:2003vv,Nepomechie:2003ez,deGier:2003iu}. The surprising fact is that this constraint involves \emph{only} the parameters which
enter the 2BTL algebra and not the coefficients in the Hamiltonian. In \cite{deGier:2005fx} it was found that these were exactly the points at which the 2BTL algebra possesses indecomposable representations.

Here we shall discuss the two-boundary problem using a special basis that we discovered for the one-boundary problem \cite{Nichols:2004fb}. We shall demonstrate the existence of special points at which there are non-trivial subspaces invariant under the action of all the 2BTL generators. They are therefore invariant under the action of the integrable Hamiltonian. We find that these are exactly the points at which Bethe Ansatz equations were written \cite{Nepomechie:2002xy,ChineseGuys,Nepomechie:2003vv,Nepomechie:2003ez,deGier:2003iu,deGier:2005fx}. Furthermore we shall show that the dimension of these invariant subspaces reproduces the splitting of eigenvalues previously obtained numerically between the two Bethe Ansatz equations. A full discussion of the Bethe Ansatz in this basis will be given elsewhere.
\section{A good basis for the one-boundary problem}
\renewcommand{\theequation}{\arabic{section}.\arabic{equation}}
\setcounter{equation}{0}  
We shall first review the bulk and one-boundary problems as these will be crucial in order to discuss the two-boundary one. For further details we refer the reader to \cite{Nichols:2004fb}.

The bulk XXZ Hamiltonian with $SU_q(2)$ quantum group symmetry is given by:
\bea
H^{TL}&=&-\sum_{i=1}^{L-1} e_i 
\eea
where the $e_i$ are the \TL generators given by:
\bea
e_i= -\half \left\{ \sigma^x_i \sigma^x_{i+1} + \sigma^y_i \sigma^y_{i+1} + \cos \gamma \sigma^z_i \sigma^z_{i+1}  - \cos \gamma + i \sin \gamma \left(\sigma^z_i - \sigma^z_{i+1} \right) \right\}
\eea
The addition of an arbitrary boundary term added to the left end is described by:
\bea
H^{1BTL}&=&-a f_0 - \sum_{i=1}^{L-1} e_i
\eea
where the parameter $a$ is arbitrary and the one-boundary \TL (1BTL) generator $f_0$ is given by \cite{Martin:1992td}:
\bea
f_0&=&-\half \f{1}{\sin(\omega+\gamma)}\left( -i \cos \omega \sigma_1^z - \sigma_1^x - \sin \omega \right) 
\eea
In \cite{Nichols:2004fb} we studied a conserved charge in this 1BTL spin chain system \cite{Delius:2001qh,Doikou:2004km}. We constructed a basis of eigenvectors of this charge which we shall refer to as the ${\bf Q}$ basis. It is defined as:
\bea \label{eqn:eigenvectors}
\left| Q_1 ; Q_2 \cdots ;  Q_L \right> &=&\left[ i e^{ - 2i \omega Q_{1}} \uparrow + \downarrow \right] \otimes \left[ i e^{ -4i \gamma Q_1 Q_{2} - 2i \omega Q_{2}} \uparrow + \downarrow \right] \nonumber \\
&& \otimes \left[ i e^{ -4i \gamma (Q_1+Q_2) Q_{3} - 2i \omega Q_{3}} \uparrow + \downarrow \right] \\
&& \otimes \left[ i e^{ -4i \gamma (Q_1+Q_2+Q_3) Q_{4} - 2i \omega Q_{4}} \uparrow + \downarrow \right] \cdots \nonumber\ \\
&& \otimes \left[ i e^{ -4i \gamma (Q_1+Q_2+\cdots+Q_{L-1}) Q_{L} - 2i \omega Q_{L}} \uparrow + \downarrow \right] \nonumber
\eea
where ${\bf Q}=(Q_1,Q_2,\cdots,Q_L)$ and $Q_i= \pm \half$.

\begin{figure}
\centering
\includegraphics[width=6 cm]{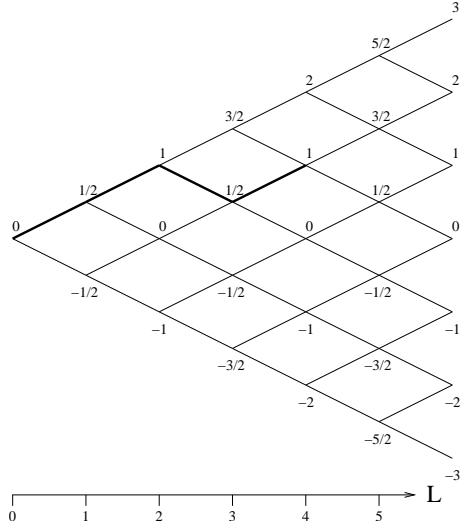}
\caption{\label{fig:fullbratelli} Full Bratelli Diagram. The system size, $L$,
  is given on the horizontal axis. The path
  corresponding to the eigenvector $\left|\half,\half,-\half,\half \right>$ is
  shown in bold.}
\end{figure}

The space of ${\bf Q}$-vectors can be encoded in a
Bratelli diagram - see figure \ref{fig:fullbratelli}.
From each different path on the diagram one reads off the values of $Q_i$ and
gets a vector from (\ref{eqn:eigenvectors}). As there are two choices at each point ($Q_i=\pm \half$) it
is obvious that this gives $2^L$ possible solutions (\ref{eqn:eigenvectors}). We shall see shortly that these solutions are not always distinct.

An important quantity is the height of a given path at point $i$. It is defined to be:
\bea \label{eqn:height}
h_i= Q_1 + \cdots +Q_i
\eea
For a system of size $L$ the degeneracy corresponding to a given value
of $h_L$ is given by:
\bea \label{eqn:Binomial}
\left(\begin{array}{c} L\\ \f{L}{2}-h_L  \end{array} \right)
\eea
In terms of the Bratelli diagram it is the
number of paths that start from the far left side at $0$ and reach that point.

One can show inductively \cite{Nichols:2004fb} that the ${\bf Q}$-basis is complete if:
\bea \label{eqn:exceptionalQ}
2 \gamma h_L +\omega \ne \pi Z
\eea
This restriction is exactly the condition that the 1BTL is non-critical. We shall assume throughout this paper that the ${\bf Q}$-basis is indeed complete. The most general case where the 1BTL algebra is also exceptional will be discussed elsewhere.

As the action for $f_0$ and $e_i$ is known in the
spin basis we can work out their action on the ${\bf Q}$-basis. For $f_0$ we have:
%
%
%
\bea \label{eqn:e0inQbasis}
f_0 \left| \half ; Q_2 \cdots ;  Q_L \right> &=& \f{\sin
  w}{\sin(\omega+\gamma)} \left| \half ; Q_2 \cdots ;  Q_L \right> \\
f_0 \left| -\half ; Q_2 \cdots ;  Q_L \right> &=& 0 \nonumber
\eea
For $e_i$ we have:
%
%
%
\bea \label{eqn:eiinQbasis}
e_i \left|  \cdots ; Q_{i-1}; \half ; \half ; Q_{i+2} ; \cdots \right> &=&0
\nonumber \\
e_i \left|  \cdots ; Q_{i-1}; \half ; -\half ; Q_{i+2} ; \cdots \right> &=&
\alpha_i \left|  \cdots ; Q_{i-1}; \half ; - \half ; Q_{i+2} ; \cdots ; \right> \nonumber
\\
&&-\alpha_i \left| \cdots ; Q_{i-1}; -\half ; \half ; Q_{i+2} ; \cdots \right>\\
e_i \left|  \cdots ; Q_{i-1}; -\half ;  \half ; Q_{i+2} ; \cdots \right> &=&
-  \beta_i \left|  \cdots ; Q_{i-1}; \half ; - \half ; Q_{i+2} ; \cdots ; \right>\nonumber
\\
&&  + \beta_i \left| \cdots ; Q_{i-1}; -\half ; \half ; Q_{i+2} ; \cdots
\right>\nonumber \\
e_i \left|  \cdots ; Q_{i-1}; -\half ; -\half ; Q_{i+2} ; \cdots \right> &=&0 \nonumber 
\eea
where:
\bea \label{eqn:AlphaandBeta}
\alpha_i&=& \f{\sin(2 \gamma h_{i-1} + \omega + \gamma)}{\sin(2 \gamma h_{i-1} + \omega)} \nonumber\\
\beta_i&=& \f{\sin(2 \gamma h_{i-1} + \omega - \gamma)}{\sin(2 \gamma h_{i-1} + \omega)}
\eea
The variables $\alpha_i$ and $\beta_i$ depend on the previous ${\bf Q}$ spins only through the height variable (\ref{eqn:height}).

In the ${\bf Q}$-basis one can see from (\ref{eqn:e0inQbasis}) and
(\ref{eqn:eiinQbasis}) that both $f_0$ and the $e_i$'s act within sectors of
a given value of $h_L=Q_1+\cdots+Q_L$. These are precisely the irreducible representations of the 1BTL algebra of size (\ref{eqn:Binomial}). The boundary generator $f_0$ is diagonalized and the bulk generators only affect nearest neighbour sites.
\section{Addition of a second boundary generator}
\renewcommand{\theequation}{\arabic{section}.\arabic{equation}}
\setcounter{equation}{0}  
We now consider the most general two-boundary XXZ model. We shall write this as:
\bea \label{eqn:2BTL}
H^{2BTL}&=& -a f_0 - a' f_L - \sum_{i=1}^{L-1} e_i
\eea
where $a$ and $a'$ are arbitrary numerical constants and the right boundary generator is given by:
\bea
f_L&=& -\f{1}{2 \sin(w_2+\gamma)} \left( i \cos w_2 \sigma^z_L + \cos \phi \sigma^x_L - \sin \phi \sigma^y_L - \sin \omega_2 \right)
\eea
This generator together with the 1BTL algebra generates the two-boundary
Temperley-Lieb (2BTL) algebra \cite{deGier:2003iu,JanReview}.

We note that the Hamiltonian (\ref{eqn:2BTL}) contains five independent boundary parameters: $\omega_1$, $\omega_2$, $\phi$, $a$, and $a'$. However only $\omega_1$, $\omega_2$, and $\phi$ are present in the boundary generators $f_0$ and $f_L$. It is  these three parameters which will control the structure of the lattice theory.

It is simple to calculate the action of $f_L$ in the ${\bf Q}$-basis:
\bea \label{eqn:RightBoundaryAction}
f_L \left|;\cdots ; Q_{L-1} ; \half \right>&=& F_{\half,\half}  \left|\cdots ; Q_{L-1} ; \half \right>+F_{\half,-\half} \left|\cdots ; Q_{L-1} ; -\half \right> \nonumber\\
f_L \left|; \cdots ; Q_{L-1} ; -\half \right>&=& F_{-\half,\half} \left|\cdots ; Q_{L-1} ; \half \right> + F_{-\half,-\half} \left|\cdots ; Q_{L-1} ; -\half \right>
\eea
where:
\bea \label{eqn:Fdefined}
F_{\half,\half}&=& -\f{\sin \left(\f{2 \gamma h_{L-1} +\omega_1-\omega_2+\phi}{2}\right)\sin \left(\f{2 \gamma h_{L-1} +\omega_1-\omega_2-\phi}{2}\right)}{\sin(\gamma+\omega_2) \sin(2 \gamma  h_{L-1}+\omega_1)}\nonumber \\
F_{\half,-\half}&=&- e^{-i(2 \gamma h_{L-1}+ \omega_1)} \f{\sin \left(\f{2 \gamma h_{L-1} + \omega_1 + \omega_2 + \phi}{2} \right)\sin \left(\f{2 \gamma h_{L-1} + \omega_1 - \omega_2 + \phi}{2} \right)}{\sin(\gamma +\omega_2) \sin(2 \gamma h_{L-1} +\omega_1)} \\
F_{-\half,\half}&=& e^{i(2 \gamma h_{L-1}+ \omega_1)} \f{\sin \left(\f{2 \gamma h_{L-1}+\omega_1-\omega_2-\phi}{2}\right)\sin \left(\f{2 \gamma h_{L-1}+\omega_1+\omega_2-\phi}{2}\right)}{\sin(\gamma +\omega_2) \sin(2 \gamma h_{L-1} +\omega_1)} \nonumber \\
F_{-\half,-\half}&=& \f{\sin \left(\f{2 \gamma h_{L-1} + \omega_1 + \omega_2 +\phi}{2} \right)\sin \left(\f{2 \gamma h_{L-1} + \omega_1 + \omega_2 -\phi}{2} \right)  }{\sin(\gamma+\omega_2) \sin(2 \gamma h_{L-1}+\omega_1)} \nonumber 
\eea
Note that $f_L$ still just affects the final site in the ${\bf Q}$-basis and only depends on the previous spins through the height (\ref{eqn:height}) at the $L-1$ site.

These expressions are well defined as we are, by assumption, away from the 1BTL exceptional points (\ref{eqn:exceptionalQ}). In the next section we shall discuss the case in which some of the $F_{\pm \half ,\pm \half}$ terms vanish.
\section{Critical points and invariant subspaces}
\renewcommand{\theequation}{\arabic{section}.\arabic{equation}}
\setcounter{equation}{0}  
If $F_{\half,-\half}$ and $F_{-\half,\half}$ are always non-vanishing (i.e. $\phi$ is generic) then there is no non-trivial invariant subspace. To prove this assume that there is an invariant subspace and take any vector within it. This will have a particular value of $h_L$. By the action of the 1BTL generators we will produce all possible vectors with the same fixed value of $h_L$. Now by the action of $f_L$ on these we will produce some vectors with $h'_L=h_L \pm 1$. Now act with the 1BTL generators on these to get all the vectors with $h_L \pm 1$. By repeating this procedure we get all $-\f{L}{2} \le h_L \le \f{L}{2}$ sectors and therefore all $2^L$ states. Therefore we conclude that there is no non-trivial subspace.

We shall now discuss the cases in which a single $F_{\pm \half,\mp \half}$ vanishes.

Let us first consider a value of $\phi$ for which there is a particular value of $h_{L-1}$, say $h_{L-1}=x$, for which $F_{\half,-\half}=0$. Now let us consider the action of the 2BTL generators on the vectors:
\bea
\left|Q_1 \cdots ; Q_{L-1} ; \half \right>
\eea
with $h_{L-1}=x$. The final boundary generator $f_L$, by the vanishing of $F_{\half,-\half}$, acts as a constant on these states. The bulk generators, except $e_{L-1}$, conserve the value of $h_{L-1}$ and therefore act completely within this space of vectors. Now the action of the generator $e_{L-1}$ on these vectors will be non-trivial only if $Q_{L-1}=-\half$. On such vectors it will give rise to vectors with $Q_{L-1}=\half, Q_{L}=-\half$ which have $h_{L-1}=x+1$. By acting with $f_L$ on these vectors we get $Q_L=\half$ states as well. Now by action of all 1BTL generators we get all vectors with $h_{L-1}=x+1$. Now we can repeat this procedure to conclude that the set of all vectors with $h_{L} \ge x + \half$ is closed under the action of all generators.  

In a similar way one can also consider a value of $\phi$ for which there is a particular value of $h_{L-1}$, say $x'$, for which $F_{-\half,\half}=0$. By a similar set of arguments we can conclude that the set of vectors with $h_{L} \le x' - \half$ is closed under the action of all generators.  

Therefore for every possible value of $h_{L-1}$ i.e. $x=-\f{L-1}{2},-\f{L-1}{2}+1,\cdots,\f{L-1}{2}$ we have a value of $\phi$ for which we get an invariant subspace. The fact that invariant subspaces only appear when parameters are tuned to particular values implies that the action of the 2BTL generators, in this representation, is becoming indecomposable at these points. The location of these critical points is exactly as previously conjectured in \cite{deGier:2005fx} found using completely different methods.

For $\phi=-2 \gamma x - \omega_1 \pm \omega_2$ (the case $F_{\half,-\half}=0$) the invariant subspace has dimension:
\bea \label{eqn:firstset}
\sum_{q \ge x + \half} \bin{L}{\f{L}{2}- q}
\eea
whereas for $\phi=2 \gamma x + \omega_1 \pm \omega_2$ (the case $F_{-\half,\half}=0$) the invariant subspace has dimension:
\bea
\sum_{q \le  x - \half} \bin{L}{\f{L}{2}- q} =  2^L - \sum_{q \ge x + \half} \bin{L}{\f{L}{2}- q}
\eea
\begin{table}[ht]
\begin{tabular}{c|cccccccccc}
Size of system & \multicolumn{10}{c}{Value of $x$} \\
& $0$ & $\f{1}{2}$ & $1$ & $\f{3}{2}$ & $2$ & $\f{5}{2}$ & $3$ & $\f{7}{2}$ & $4$ & $\f{9}{2}$ \\
\hline
1 & 1 \\
2 & - & 1 \\
3 & 4 & - & 1 & \\
4 & - & 5 & - & 1 \\
5 & 16 & - & 6 & - & 1 \\
6 & - & 22 & - & 7 & - & 1 \\
7 & 64 & - & 29 & - & 8 & - & 1 \\
8 & - & 93 & - & 37 & - & 9 & - & 1 \\
9 & 256 & - & 130 & - & 46 & - & 10 & - & 1\\
10 & - & 386 & - & 176 & - & 56 & - &  11 & - & 1
\end{tabular}
\caption{\label{tab:2BTLcriticalpts} Dimension of the invariant subspaces at the 2BTL critical points.}
\end{table}
In table \ref{tab:2BTLcriticalpts} we give the values of (\ref{eqn:firstset}) for $x \ge 0$. The points at which the non-trivial subspaces exist are exactly the points at which the Bethe Ansatz can be performed \cite{ChineseGuys,Nepomechie:2002xy,Nepomechie:2003vv,Nepomechie:2003ez}. At these points the eigenvalues of the two-boundary Hamiltonian (\ref{eqn:2BTL}) split into two sets. These sets are each described by Bethe Ansatz equations. This is similar to the diagonal chain where two sets of Bethe Ansatz equations come from the two distinct Bethe reference states. Here we only wish to draw attention to the fact that in several cases the dimensions of the invariant subspaces give exactly the numerically observed splitting.

In \cite{Nepomechie:2003ez} it was found in the $x=\half$ case that the number of solutions of one of the Bethe Ansatz equations followed the formula:%
\bea
2^{L-1} + \half \left( \begin{array}{c} L \\ \f{L}{2} \end{array} \right)
\eea
The other eigenvalues were given as solutions to the other Bethe Ansatz equation. As there are in total $2^L$ eigenvalues these are therefore of number:
\bea
2^{L-1} - \half \left( \begin{array}{c} L \\ \f{L}{2} \end{array} \right)
\eea
Evaluating this for $L=2,4,6,8,10$ we find $1,5,22,93,386$ which are exactly the numbers in the $x=\half$ column of table \ref{tab:2BTLcriticalpts}.

In the case of $L$ odd and $x=0$ it was found numerically in \cite{Nepomechie:2003ez} that each Bethe Ansatz contained $2^{L-1}$ solutions. Again this is in agreement with table \ref{tab:2BTLcriticalpts}. 

We therefore conjecture that the numbers given in the table above correctly account for the splitting of the $2^L$ solutions between the Bethe Ansatz equations (at least when parameters other than $\phi$ are generic). The numbers in the table correspond to the size of the smaller set.

A proper explanation of this fact requires the Bethe Ansatz equations to be formulated directly in the ${\bf Q}$-basis. This will be discussed in detail elsewhere.

We would finally like to point out that although we have used the one-boundary ${\bf Q}$-basis for the left-hand boundary it is equally possible to consider the two-boundary problem by beginning at the right-hand side.
\section{Conclusion}
We have discussed the XXZ model with general non-diagonal boundary terms. We first isolated the algebraic aspects of the problem by rewriting the model in terms of the 2BTL algebra. By calculating the action of the 2BTL generators on a good basis for the one-boundary problem we were able to show that at particular points there are non-trivial invariant subspaces. The dimension of these subspaces was also calculated. The fact that the Bethe Ansatz was only able to be written at these critical points, and moreover that the number of solutions to each Bethe Ansatz equation is given precisely by the dimension of the invariant subspace, still needs to be properly explained.

We believe that it is possible to generalize the ${\bf Q}$-basis to study more general spin chains with boundaries. We shall return to this point at a later date.
\section*{Acknowledgements}
This research is partially supported by the EU network
\emph{Integrable models and applications: from strings to condensed matter} HPRN-CT-2002-00325 (EUCLID). 

This present paper arose, partially, from a more detailed mathematical investigation of the 2BTL algebra in collaboration with B. Westbury. I would also like to thank J. de Gier, P. Pyatov, and V. Rittenberg for useful discussions on the 2BTL algebra.
%

\end{document}